\documentclass[pdflatex,sn-mathphys]{sn-jnl}

\jyear{2021}%

\theoremstyle{thmstyleone}%
\theoremstyle{thmstyletwo}%

\theoremstyle{thmstylethree}%

\begin{document}

\title[A Container-Based Workflow for Deep Learning in HPC]{A Container-Based Workflow for Distributed Training of Deep Learning Algorithms in HPC Clusters}


\author*[1]{\fnm{Jose} \sur{Gonz{\'a}lez-Abad}}\email{gonzabad@ifca.unican.es}

\author[1]{\fnm{{\'A}lvaro} \sur{L{\'o}pez Garc{\'i}a}}\email{aloga@ifca.unican.es}
\equalcont{These authors contributed equally to this work.}

\author[2]{\fnm{Valentin Y.} \sur{Kozlov}}\email{valentin.kozlov@kit.edu}
\equalcont{These authors contributed equally to this work.}

\affil*[1]{\orgname{Instituto de F{\'i}sica de Cantabria (IFCA), CSIC-Universidad de Cantabria}, \orgaddress{\city{Santander}, \country{Spain}}}

\affil[2]{\orgname{Karlsruhe Institute of Technology (KIT), Steinbuch Centre for Computing (SCC)}, \orgaddress{\city{Karlsruhe}, \country{Germany}}}


\abstract{Deep learning has been postulated as a solution for numerous problems in different branches of science. Given the resource-intensive nature of these models, they often need to be executed on specialized hardware such graphical processing units (GPUs) in a distributed manner. In the academic field, researchers get access to this kind of resources through High Performance Computing (HPC) clusters. This kind of infrastructures make the training of these models difficult due to their multi-user nature and limited user permission. In addition, different HPC clusters may possess different peculiarities that can entangle the research cycle (e.g., libraries dependencies). In this paper we develop a workflow and methodology for the distributed training of deep learning models in HPC clusters which provides researchers with a series of novel advantages. It relies on udocker as containerization tool and on Horovod as library for the distribution of the models across multiple GPUs. udocker does not need any special permission, allowing researchers to run the entire workflow without relying on any administrator. Horovod ensures the efficient distribution of the training independently of the deep learning framework used. Additionally, due to containerization and specific features of the workflow, it provides researchers with a cluster-agnostic way of running their models. The experiments carried out show that the workflow offers good scalability in the distributed training of the models and that it easily adapts to different clusters.}

\keywords{Distributed Training, Deep Learning, High Performance Computing, udocker, Docker, Horovod}



\maketitle

\section*{Declarations}


\begin{itemize}
\item Funding: This work has been supported by the project DEEP-Hybrid-DataCloud ``Designing and Enabling E-infrastructures for intensive Processing in a Hybrid DataCloud'' that has received funding from the European Union's Horizon 2020 research and innovation programme under grant agreement number 777435. J. Gonz{\'a}lez-Abad and A. L{\'o}pez Garc{\'i}a would also like to acknowledge the support of the funding from the Spanish \textit{Agencia Estatal de Investigación} through the \textit{Unidad de Excelencia María de Maeztu} with reference MDM-2017-0765. A part of this work was performed on the supercomputer ForHLR funded by the Ministry of Science, Research and the Arts Baden-Württemberg and by the Federal Ministry of Education and Research.
\item Conflict of interest/Competing interests (check journal-specific guidelines for which heading to use): Not applicable
\item Ethics approval: Not applicable
\item Consent to participate: Not applicable
\item Consent for publication: Not applicable
\item Availability of data and materials: Data and materials are available, URLs are provided through the manuscript
\item Code availability: Code is available, URLs are provided through the manuscript
\item Authors' contributions: All authors contributed equally to the study conception and design of the workflow. All authors read and approved the final manuscript.
\end{itemize}

\section{Introduction}
\label{intro}

The machine learning subfield known as deep learning \cite{goodfellow_deep_2016} has been the protagonist of a revolution in recent decades. The growing amount of data available as well as the increase in the computing power of current hardware has allowed neural networks algorithms to become state-of-the-art (SOTA) in multiple complex tasks such as language processing, computer vision and image generation. These successes have led researchers in more specific scientific areas to begin applying these techniques in their own fields. For this reason, these algorithms can now be seen applied in fields as diverse as particle physics \cite{baldi2014searching,de2016jet}, cosmology \cite{tuccillo2018deep,primack2018deep}, biology \cite{jumper2021highly} or climatology \cite{scher2018toward,rasp2018deep}, just to cite some examples.

Given their nature, this kind of algorithm is very resource-intensive. The most recent SOTA results have been obtained through large clusters, which are generally provided with specialized hardware targeted to accelerate the execution of these algorithms. Specifically, in deep learning, graphical processing units (GPUs) have shown the greatest speedup. This has led to the development of powerful clusters equipped with multiple GPUs communicating between each other. Usually, researchers get access to these machines in order to run their experiments, but installing and configuring the required software to train a deep learning model in these multi-user environments is not an easy task, mainly due to restrictions in permissions. Also, the deployment of models in such distributed environments requires advance knowledge in computing, which applied scientists may not have. There are cloud technologies with simple interfaces that seek to reduce this barrier \cite{lopez_garcia_cloud-based_2020}, but due to particular requirements of academia and research (e.g. high throughput and low latency networks) it is still common to access High Performance Computing (HPC) clusters.

One way to alleviate these problems is by allowing users of HPC clusters to use containerization tools to encapsulate and run specific software stacks. This kind of tools allows the isolation of software from the rest of the system through the use of specific features of the Linux kernel (e.g., kernel namespaces and Linux controls groups). These solutions are very popular and have been widely adopted in both industry and academia. In the former, they have become an important part of the development and deployment of applications and services, while in the latter they have helped increase the portability of the software required to run an experiment, as well as ease its reproducibility. 

The most widespread containerization tool is Docker \cite{merkel_docker_2014}, however, given the escalation of privileges needed, it is not suitable for multi-user environments such as HPC clusters. Due to this, new containerization tools focused on running in such environments began to be developed, among which Singularity \cite{kurtzer_singularity_2017}, Charliecloud \cite{priedhorsky_charliecloud_2017} and Shifter \cite{gerhardt_shifter_2017} stand out. Some of these tools, while allowing non-root users to execute their containers, still need root permissions for the proper installation and configuration of the tool. Some others avoid this by recurring to specific features of the Linux kernel.

The tool used in the experiments described in this paper is udocker \cite{gomes_enabling_2018}. It is a user tool oriented to the execution of Docker containers without the need for root permissions or any other kind of privileges. This tool supports a large subset of Docker commands, so the researcher only needs to know a few notions of the latter. Given udocker's sole dependency on Python, the researcher hardly needs any administrator intervention for installing it, in contrast with other solutions. In this way, users may safely leverage new libraries and frameworks without recurring to HPC administrators, making the research cycle more dynamic.

Running scientific experiments as containers in an HPC cluster is not trivial. Sometimes the code needs to be adapted and special versions of software may need to be installed in order to take advantage of specialized hardware. For this reason, several workflows have recently been developed \cite{younge_tale_2017,brayford_deploying_2019,brayford_deploying_2020} to combine in a single flow the entire process required to run an experiment, from the development of the experiment to its deployment on a cluster, using the previously mentioned containerization tools.

In this paper, we develop a container-based workflow to train deep learning models in multiGPU environments in HPC clusters. To the best of our knowledge, this workflow is the first one specifically oriented to train this kind of models on specialized hardware in a distributed manner while filling previous research gaps. This workflow provides HPC users with a series of novel advantages:

\begin{itemize}
  \item No administrator or superuser intervention is needed, accelerating the research cycle.
  \item Cluster-agnosticism, which allows researchers to run their experiments in any x86-64 architecture cluster available with minor changes.
  \item Efficient scaling of the training of the model with respect to native setups.
\end{itemize}

For this, we use udocker as containerization tool and Horovod \cite{sergeev_horovod_2018} as library for distributing the training of the model across GPUs. Its use allows the distribution of the training independently of the framework used (TensorFlow, Keras, Pytorch and MXNet). The conjunction of these tools, developed integration efforts and good practices conform the workflow presented in this work. As a result, we contribute with a novel workflow built on the difficulties that researchers face when training deep learning models on multiGPU HPC clusters. Furthermore, we show these advantages by running different synthetic and real experiments on different clusters, without requiring administrator intervention.

The paper is structured as follows. Section \ref{sec:1} introduces all the technologies involved in the workflow. Section \ref{sec:2} introduces related work and shows how our workflow fills the gaps in the literature. In Section \ref{sec:3} the actual workflow is presented, while Section \ref{sec:4} describes the experiments.  Section \ref{sec:5} details the systems where these experiments were executed, and Section \ref{sec:6} shows their results. Finally, Section \ref{sec:7} contains the discussion of these results and Section \ref{sec:8} the conclusions of the work. 

\section{Background}
\label{sec:1}

\subsection{Distributed Deep Learning}
\label{sec:1:1}

A GPU used to be enough to train a deep learning model. It allowed researchers to speed up the training process and therefore the experimentation with respect to a standard central processing unit (CPU). Due to the increase in the complexity of these kinds of models and the datasets used to train them, environments with just a GPU started to introduce limitations caused by its low available memory and increased training time.

As a consequence of these complexities, researchers had to start using distributed infrastructures to train their models. These infrastructures are usually made up of several computing nodes, each of which is equipped with one or more GPUs. This heterogeneity brings several distributed computing concepts to take into account, such as communication between nodes, storage management and effective use of computing resources. This fact entangles researchers' work, since these infrastructures are often necessary to achieve SOTA in deep learning \cite{vaswani_attention_2017,karras_analyzing_2020}.

The two most commonly used parallelization strategies for this kind of models are data parallelism and model parallelism \cite{nguyen_machine_2019}. In the former, each worker (GPU in this case) loads a copy of the model and computes the gradients of a chunk of training data. These gradients are then averaged and used to update each of the models. The difficulty of this strategy lies in the synchronization of the parameters between workers. In the latter, the model is split between workers, so each of them computes the correspondent part of the forward and backward pass. This split is not trivial and is determined by the model itself. Considering its flexibility and ease of use regarding model parallelism, data parallelism is more widely used by researchers.

There are many deep learning frameworks available, each with its characteristics and syntax \cite{nguyen_machine_2019}.  Based on the framework, the parallelization strategy may vary, adding another layer of complexity to the experimentation process. Tensorflow \cite{abadi_tensorflow_2016} supports both kinds of strategies, focusing on data parallelism. Depending on the selected strategy, it is possible to use different synchronizations between workers (parameter server and all-reduce algorithms). All these strategies are also accessible within Keras \cite{chollet2015keras} through the \texttt{tf.Keras} API. In the same way, Pytorch \cite{paszke_pytorch_2019} offers a variety of strategies based on data parallelism in addition to some strategies focused on more specific scenarios like RPC Distributed Training.

The variety of deep learning frameworks and their implementations of distributed strategies complicates the task of researchers without advanced knowledge in distributed systems since they must first know which strategy is the best for them in terms of efficiency and scalability with respect to their model, know extremely well the syntax of the framework they are using and properly adapt the code for the distributed execution. As a result of the need to simplify this process, Horovod was born.

\subsubsection{Horovod}
\label{sec:1:1:1}

Advances in deep learning made large companies like Uber find it necessary to scale their models to multiGPU environments. They began by testing Tensorflow, the most widespread framework at that time, but they ran into some issues. The first was its steep learning curve due to the high number of concepts necessary for the correct implementation, the second was the poor scalability achieved. These facts led Uber to develop Horovod \cite{sergeev_horovod_2018}, a distributed deep learning training framework built over deep learning frameworks like Tensorflow, Keras and Pytorch. Twin pillars of Horovod are:

\begin{itemize}
  \item \textbf{Ease of use}: Regardless of the framework we use, with Horovod we only need to add a few lines of code to 
     train our model in both GPU and multiGPU environments. This removes barriers derived from differences between syntax.
  \item \textbf{Scalability}: Its use ensures good scalability thanks to its implementation based on various communications
     standards \cite{sergeev_horovod_2018}.
\end{itemize}

All communication between workers is done using a collective communications library. Horovod allows the use of either a Message Passing Interface (MPI) implementation like OpenMPI or Gloo, a Facebook library oriented to communicate workers in machine learning applications. These libraries are in charge of the synchronization of the parallel processes and allow the user to correctly manage the execution.

The distribution of the training could be done with one of these libraries, but operations like gradients average among workers would introduce bottlenecks in the process. Horovod recently added the NVIDIA Collective Communication Library (NCCL) to its framework. This library includes communication routines optimized for NVIDIA GPUs that speed up data exchange. Among those included, the ring-Allreduce based on an implementation inspired by \cite{patarasuk_bandwidth_2009} stands out. In this algorithm, each worker communicates with two others. During these communications, the workers exchange data and perform reduce-like operations. This process is repeated until the complete reduction of the data.

At present, Horovod combines MPI or Gloo for CPU oriented operations and process management and NCCL for GPU communications. All these features make Horovod an appropriate library for the distributed training of deep learning algorithms.

\subsection{HPC Clusters}
\label{sec:1:2}

In research and academia, scientists normally gain access to GPUs through HPC clusters. These are usually managed by computing centers specialized in this type of technology that offers their computing power as a service to a wide range of researchers.

These computing clusters, which typically run under a Linux operating
system\footnote{\url{https://www.top500.org/statistics/details/osfam/1/}}, are composed of a collection of different nodes connected via high throughput and low latency interconnects like Infiniband\footnote{\url{https://www.infinibandta.org/}}. In this way, it is possible to parallelize tasks between nodes and process large amounts of data efficiently. In some of these clusters, several of their nodes are provided with a GPU besides the CPU. Through access to several of these nodes, it is possible to distribute the training of deep learning models.

Once researchers have gained access to an HPC environment, they connect remotely and execute their workloads through a batch system or job scheduler such as Slurm \cite{yoo_slurm_2003}. For the proper execution of the job, it is necessary to correctly configure the environment in which it will run. This mainly includes the configuration and installation of programming tools and libraries. In jobs that need to be run on GPUs, this task becomes even more complicated since they depend on more complex applications linked to the GPU like the Compute Unified Device Architecture (CUDA) or the NVIDIA CUDA Deep Neural Network library (cudNN). Because deep learning frameworks depend on these applications, their installation must be done taking into account the versions of these available in the HPC cluster. 

\subsection{Containerization technologies}
\label{sec:1:3}

When a new user accesses an HPC facility, the first thing they need to do is set up their own environment. In these systems user permissions are restricted and the computing environment (i.e., operating system and libraries) is fixed, which can lead to issues in basic aspects like software management (e.g., no package manager available) or no administrative privileges, limiting the usage of certain software. These problems are common in computational science and are present when the user is forced with a fixed environment \cite{oesterle_experiences_2015}. Moreover, once the setup is done at one facility, it may not correctly adapt to the heterogeneity of systems that researchers can manage, since if they want to deploy the same environment in another HPC they need to redo the installation with the difficult additions of doing it on a different system. They may also find differences between the local system on which they programmed the application and the HPC on which they deploy it. These issues add another layer of complexity to the research process.

One solution to these problems is containerization. This technology allows all the software, settings, code and environment variables of a user to be packaged in what is known as containers, then this software stack can be deployed on any machine with ease. Unlike virtual machines that virtualize at the hardware level through a hypervisor, containers virtualize at the operating system (OS) level using several features of the Linux kernel. Containers are isolated processes running on top of the OS kernel, so there is no need for hardware virtualization. These features make containers more efficient than Virtual Machines \cite{felter_updated_2015}.

Specifically, in HPC environments, the use of containers can make the development and deployment process easier for researchers. First, they can stack their environment in an isolated file, which allows them to take better control of aspects like the libraries used and their versions. This stack allows them to run the same both in their local environment and in HPC ones, avoiding tedious installations. In addition, it also facilitates reproducibility since this file can include everything necessary to run these experiments (e.g., code, models, seeds). The latter plays a fundamental role in the field of deep learning, where the exact reproducibility of the results does not depend solely on the code. We briefly present some of the most popular containerization technologies in what follows.

\subsubsection{Docker}
\label{sec:1:3:1}

Docker \cite{merkel_docker_2014} is the most popular container creation and execution solution. Its implementation is based on two features of the Linux kernel: \texttt{namespaces} and \texttt{cgroups}. The former is responsible for building the layer of isolation, making the container processes run isolated from the rest of the host. The latter is in charge of managing and limiting the hardware resources assigned to the containers. In addition, Docker also allows the distribution of containers thanks to DockerHub, a registry service focused on sharing container images. Unfortunately, this solution has a major drawback in our context, which is the dependency on a root owned daemon process, requiring, therefore, administrative privileges to run a container. In multi-user systems like HPC clusters, a standard user does not have these privileges, making Docker unsuitable for these types of environments.

\subsubsection{Singularity}
\label{sec:1:3:2}

Singularity \cite{kurtzer_singularity_2017} is a tool that offers image-based containers which allow software encapsulation and its execution in HPC environments, given its non-root execution. It is based upon \texttt{setuid} file permission, which allows running a program with escalated permissions. A \texttt{setuid} binary gives root access to certain basic operations required for the proper execution of the container. This method is flexible but dangerous from a security perspective. It is also possible to run Singularity in a \texttt{non-setuid} mode, but it requires unprivileged user namespaces, a fairly recent Linux kernel feature that may not be available on all HPC environments. Moreover, apart from the security implications, the main issue with \texttt{setuid-mode} is that the binary needs to be previously installed with root permissions, so the intervention of an administrator to install and configure this tool is needed beforehand. 

\subsubsection{udocker}
\label{sec:1:3:4}

udocker \cite{gomes_enabling_2018} is a tool that enables the execution of containers without requiring administrative privileges. It focuses on the execution of containers in multi-user environments, leaving the process of creation and distribution of images to Docker. In
order to run these Docker containers, it is necessary to implement chroot-like functionality without the need for root access. This is achieved in three different ways: unprivileged user namespaces, \texttt{PTRACE} mechanism, and \texttt{LD\_PRELOAD} mechanism. The first works the same as in Singularity. The second allows changing the pathnames of the system calls in runtime, making them reference pathnames within the container. The third allows the overriding of shared libraries, which allows running those of the chroot-environment (see \cite{gomes_enabling_2018} for more details). The required binaries for these operations are statically compiled, which increases portability. udocker does not require any restrictive permissions beyond the user's own, eliminating the need for administrative privileges required by other tools for its installation and execution. This fact also provides more autonomy to users, since they do not need to rely on the administrator's intervention.

\section{Related Work}
\label{sec:2}

The need to run experiments in multi-user environments has led to the development of workflows oriented to such purpose. In \cite{lopez_garcia_cloud-based_2020} a set of tools and services to cover and ease the whole machine learning development cycle is presented, although useful it focuses on cloud computing and not on HPC clusters. Regarding these environments, the increased efficiency and lack of overhead of containers compared to virtual machines \cite{torrez_hpc_2019} has led to the development of container-based workflows.

In \cite{chung_using_2016} the use of Docker containers in HPC clusters is studied. They compare the deployment of applications with Docker containers and virtual machines and conclude that the former has more advantages in addition to no overheads. In \cite{sparks_enabling_2019} authors extend the Docker platform to meet the requirements of HPC systems and ease the execution of workflows on such environments. In \cite{azab_enabling_2017} a Docker's wrapper is developed to execute containers via the Slurm Workload Manager in the cluster of the University of Oslo. These workflows rely on Docker, so it is still necessary for the cluster to already have this system installed or for the user to ask the administrator to install it. This dependency makes these workflows non-cluster-agnostic, since the user can not extend them to other clusters without the respective administrators' intervention.

In parallel, researchers developed workflows with the same objectives but using containerization tools focused on HPC environments. In \cite{younge_tale_2017} a workflow based in Singularity is proposed for the deployment of containers developed in a local environment both to the cloud and to an HPC cluster. In \cite{jacobsen_contain_2015} authors develop a workflow that allows users to run containers in a cluster using Shifter \cite{gerhardt_shifter_2017}, a software oriented to running containers in multi-user environments. Although these workflows are well adapted to multi-user environments, they still depend on an administrator since Singularity needs a daemon with root permissions and Shifter needs to be installed on all nodes where the experiment is run. Furthermore, these works still focus on a single cluster.

The developments presented so far are for general purpose experiments. Deep learning experiments have their own characteristics that make them even more complex to run in HPC contexts (e.g., requiring specific libraries to control GPUs). In \cite{brayford_deploying_2019} authors focus on applications oriented to deep learning, for which they use Charliecloud \cite{priedhorsky_charliecloud_2017}. Subsequently, a similar workflow is applied to the training of a deep learning model in the field of particle physics \cite{brayford_deploying_2020}. The developed workflow allows them to take advantage of the hundreds of CPUs available in their cluster. These workflows, although they focus on deep learning algorithms, do not use specialized hardware (GPUs) which usually requires more complex configurations. Furthermore, many of these workflows are still focused on a specific cluster, therefore not being cluster-agnostic.

In \cite{grupp_benchmarking_2019} a study of the scalability of udocker with respect to the number of GPUs used to train a deep learning model is presented. udocker performance is compared against the native environment and other containerization tools, and it is observed that it does not present a significant overhead in multiGPU scenarios. 

In addition to the containerization tools explored in Section \ref{sec:1:3}, there are others that seek to fulfill the same purposes. Podman \cite{gantikow_rootless_2020} is another containerization tool with features similar to those of Docker, which also offers the possibility of running containers without requiring root permissions. The disadvantage of such tools with respect to the developed workflow is that they either require administrator permissions for installation or, for compilation, depend on software that may not be available in all clusters (e.g. Podman requires Go\footnote{\url{https://go.dev/}} for building it from source). udocker installation can be performed by the user and its main dependency is Python, the current most popular programming language. Other tools aim to facilitate the management and installation of libraries in multi-user spaces, for example Spack \cite{gamblin_spack_2015}, a package manager oriented to these environments. Through a simple syntax, it allows users a better management of versions and configurations options in HPC clusters. However, it does not have all the benefits that a containerization tool can offer in the context of this work (see \ref{sec:1:3}).

Recently, work has been developed that explores the future difficulties that researchers will face when scaling up their experiments on HPC clusters. In \cite{hob2020enabling} the difficulties involved in moving workflows from CPU-based to GPU-based systems are addressed. Authors propose initial steps towards a framework for the automatic deployment of optimized containers in such systems. In \cite{canon2019case} authors provide an analysis of the challenges in providing portability and reproducibility to containers in HPC. One of these challenges is the need of mapping specific host libraries to the container. 

In this work we develop a new workflow offering a number of advantages over more conventional ones. It is based on udocker, so the user does not need administrator intervention to run their containers, what is more, this allows them to run their containers in a cluster-agnostic way since users can manage its installation relying on a very basic set of dependencies. The independence that this workflow gives to the user (no administrators' intervention required) allows for an increase in the speed of the research cycle. Previous work has explored the performance of udocker, and it has been seen that scalability is not harmed, thus being able to easily scale through various GPUs. All this makes udocker the perfect candidate to develop workflows oriented to the training of deep learning models in multi-user multiGPU environments in a cluster-agnostic way. 

\section{Methodology and Workflow}
\label{sec:3}

\begin{figure}[ht]
    \centering
    \includegraphics[width=\linewidth]{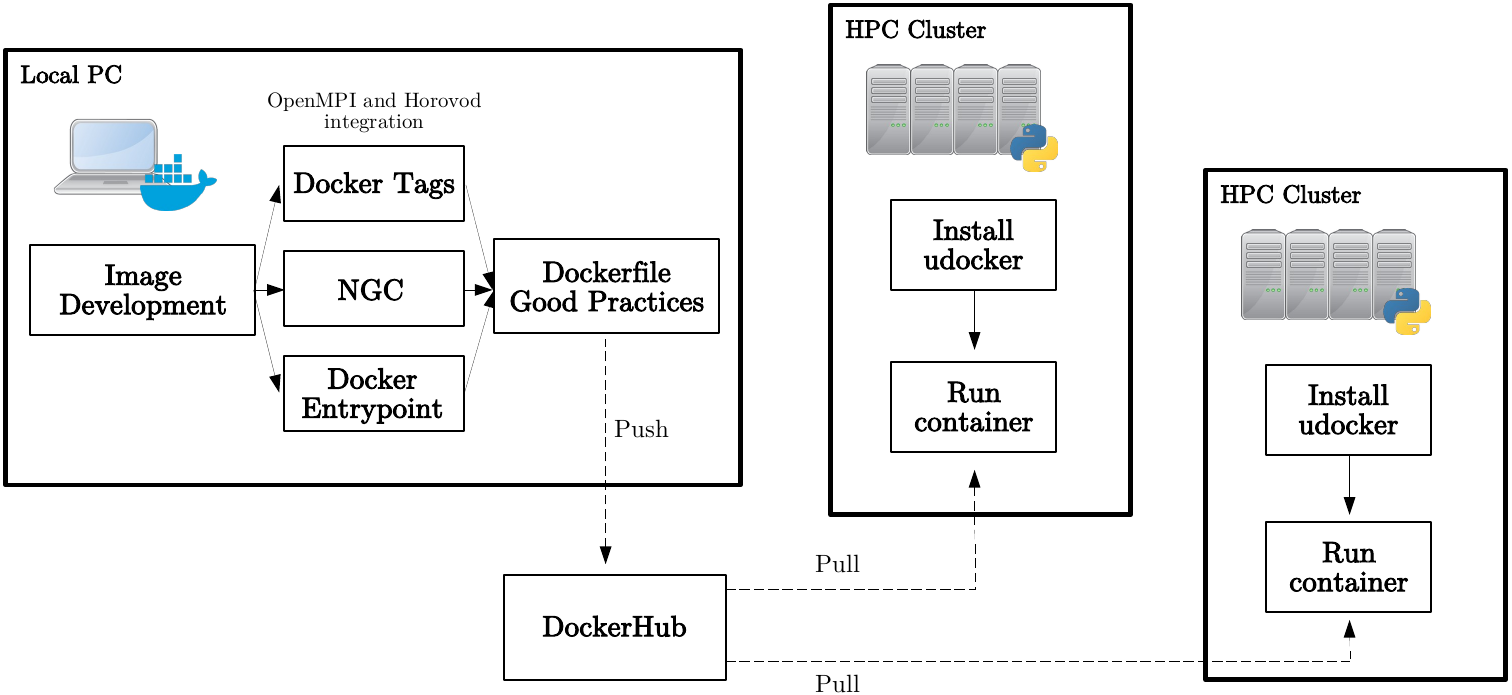}
    \caption{Diagram of the developed workflow}
    \label{fig:1}
\end{figure}

Given the aforementioned necessity of training deep learning models in distributed environments, we have developed a methodology focused on this task when some HPC cluster is involved. It is based mainly on the interaction between three tools: Docker, udocker and Horovod. Figure \ref{fig:1} shows a diagram of the workflow.

Under this methodology, researchers develop the image of their local environment with Docker. They can start from images already built and tested by the community to support computation on GPUs. Once they have an image of their experiment's environment, they upload it to DockerHub making it accessible through the Internet. When they want to deploy this image in any HPC cluster, they just need to install udocker in the cluster and download the image from DockerHub. Horovod is in charge of distributing the model training among the selected nodes of the cluster without harming scalability.

In this section, we explore this workflow and study the different problems that arise and how to tackle them.

\subsection{Image Development}
\label{sec:3:1}

Docker provides virtualization through running images as containers. An image is a file that contains all the code, libraries, tools and dependencies that define the environment. A container is a virtualized runtime environment responsible for isolating the execution of the image from the rest of the environment. The image building process is defined in a file known as Dockerfile.

A Dockerfile is a file belonging to the Docker environment in which all the commands used to build the image are defined as a series of steps. When the image is built, these commands are executed sequentially. Each command generates a layer that is  stacked on top of the ones generated with previous commands. In this way, the final image is conformed of a set of read-only layers. When a container of a certain image is executed, Docker stacks a new read-write layer that allows to work on top of the previously installed ones.

Setting up working environments with GPU support is not a trivial task. Correct installation and configuration of GPU-related libraries such as CUDA, cudNN and NCCL are required. This task is complicated in most HPC environments given restrictions on user permissions and dependency on already installed software. Thanks to the Docker's layer structure and the available already configured images (e.g., TensorFlow, NVIDIA), it is possible to avoid this tedious configuration, since we can stack custom layers over those already supporting the needed GPU libraries. 

Horovod allows researchers to abstract from the computational complexities involved in the adaptation to training in distributed environments. Its use does not limit researchers in any way, since it can be used with the most popular deep learning frameworks with minor changes in the syntax, which allows the standardization of the training with respect to the framework used.

Once this basic software is installed, researchers can move on to installing more specific things, such as the libraries or additional programming languages that their experiments require. During this process they can define the versions of these, ensuring an environment fully compatible with their experiment. 

This process can be carried out in its entirety in the local environment, which allows researchers to test the developed environment easily and safely. Docker also has instructions like \texttt{COPY} that allow copying local files to the image. This allows researchers to include code in the image so that not only they can have their entire environment in a single file, but also the code to reproduce their experiment. In this way, researchers can easily develop a fully reproducible experiment while working on it. This removes well-known problems as libraries version mismatches or compatibility issues, since all the elements required to reproduce the experiment are encapsulated in the same environment.

Once researchers have defined the Dockerfile, they own a file that contains the specific environment of their research, from OS to high level libraries passing through specific dependencies and GPU-related software. For example, a general image with GPU-support, OpenMPI, and Horovod could be built and shared among researchers. Taking this image as a basis, other researchers would only need to add their code and specialized libraries to train deep learning models in any HPC cluster under the workflow developed in this work.

\subsection{OpenMPI and Horovod integration}
\label{sec:3:2}

Parallel computing systems such as HPC clusters need message passing approaches to communicate the different nodes. MPI is the de facto standard when it comes to message passing. There are many implementations of this standard, OpenMPI \cite{gabriel04:_open_mpi} being the most
widespread. OpenMPI is an open-source MPI implementation that enables efficient communications between applications on different nodes and supports advanced protocols such as Infiniband.

Horovod coordinates work between processes via MPI, using OpenMPI as the basis for \texttt{horovodrun}, its MPI wrapper. In addition, MPI is also needed to communicate the different udocker containers distributed among nodes. Due to inconsistencies between some commands, it is necessary that the version of OpenMPI available in the cluster and that of the udocker container match. This fact introduces difficulties in the workflow, since now researchers depend on the version of OpenMPI installed in the cluster in which they run the experiment, making the process non-cluster-agnostic. We propose several solutions to this issue. 

\subsubsection{Docker Tags}
\label{sec:3:2:1}

The most straightforward solution would be to build the image taking into account the version of OpenMPI available in the HPC in which we want to run the experiment. The main problem is that this solution complicates the adaptation of the workflow to various HPC environments.

One way to tackle this problem is by using Docker tags. Tags are names used to group an image with respect to, for example, some version or variant. In this case, researchers could, from their base image, upload variants to DockerHub according to the OpenMPI version installed and classify them based on tags. Hence, they can choose which tag to pull based on the OpenMPI version available in the HPC cluster in which they plan to run the experiment.

\subsubsection{NGC catalog}
\label{sec:3:2:2}

NVIDIA GPU Cloud\footnote{\url{https://catalog.ngc.nvidia.com/}} (NGC) is a catalog of GPU optimized software developed by NVIDIA. It offers images already built and optimized for the training of deep learning models in multiGPU environments. Its objective is to provide researchers with ready-to-run images which they can pull in any environment, thus avoiding complex installations and configurations.

The images it offers are optimized and support the latest NVIDIA hardware. They can be classified based on the included software and applications. It also offers software development kits (SDKs) for GPU computing in HPC clusters. Some of these images come with Horovod and OpenMPI integrated, so they can be used to train models in multiGPU environments within our developed workflow as long as these versions match the ones available in the HPC cluster.

NVIDIA offers this as already built images, so researchers have less control over the software installed on them. This can lead to incompatibilities between the included software and the one researchers need to install, making proper configuration difficult. Furthermore, the software included in this catalog is mainly optimized for GPUs with an architecture equal to or more recent than Pascal.

\subsubsection{Docker Entrypoint}
\label{sec:3:2:3}

\begin{figure}[ht]
    \centering
    \includegraphics[scale=0.11]{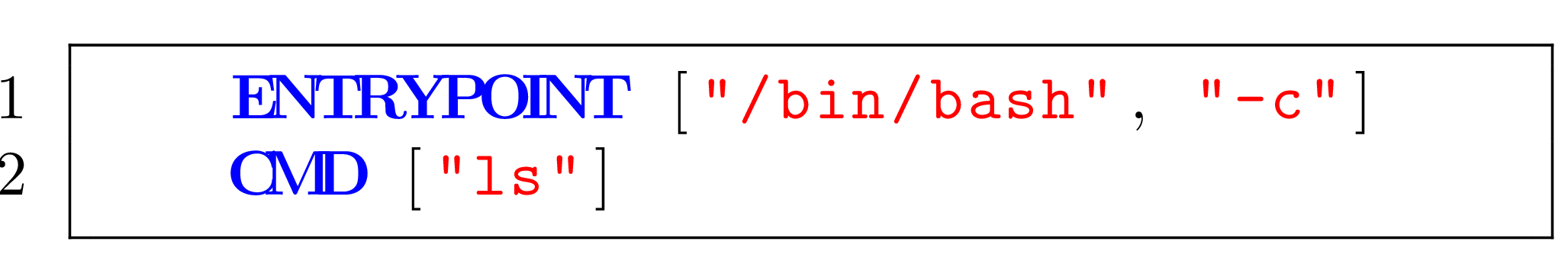}
    \caption{Dockerfile's code to run a command by default when the container is executed}
    \label{fig:2}
\end{figure}

When a container is executed, Docker runs a command that can be defined in the Dockerfile through the \texttt{ENTRYPOINT} instruction. This command is executed whenever the container is run. It is possible to pass arguments to this command through the \texttt{CMD} instruction. These parameters can be overwritten when we run the container, an example is shown in Fig.\ref{fig:2}.

In this case, the instruction \texttt{docker run <image>} makes the container run the \texttt{ls} command. It is possible
to overwrite what we pass to the entrypoint with \texttt{docker run <image> <cmd>}. Taking this into account, researchers could,
for example, execute a certain script when the container is started.

Based on this, we have developed a script that installs OpenMPI and Horovod as follows:

\begin{enumerate}
    \item Takes the versions of OpenMPI and Horovod to install provided by the user as arguments.
    \item If they are already installed, skip the process, otherwise proceed with the installation.
    \item Execute the command entered by the user as an argument (if none is entered, execute \texttt{/bin/bash}).
\end{enumerate}

It is possible to include this script as entrypoint in the Dockerfile so when researchers run the container it installs the OpenMPI and Horovod versions requested by them, when the installation finishes the container tool (e.g., Docker, udocker) executes the command passed as
\texttt{cmd} argument. This allows having an image without OpenMPI and Horovod installed, which gives a lot of flexibility  since researchers can define their image without taking into account the OpenMPI version available in the cluster, making their image cluster-agnostic. When researchers start working in a specific cluster, they just need to run the container, taking into account the version of OpenMPI available there. The first time they run the container it installs the versions required by the cluster, later executions do not need to install them. The installation of Horovod is needed given its dependency on OpenMPI version.

In this section, we offered three alternatives to solve the dependency problem between OpenMPI versions in the image and the cluster. Tags are a simple solution, although it can complicate the researcher's workflow since they have to manage multiple images very similar to each other. The NGC catalog offers images ready to run this type of applications but limits the user to the software already installed in them. The entrypoint approach seems to be the best option given its flexibility, since the user ends up having a single image that can be executed in any cluster regardless of the OpenMPI version installed in it. 

\subsection{DockerHub}
\label{sec:3:3}

In addition to offering tools for building images, Docker also offers DockerHub, a hosted repository service that enables their distribution. Through this, it is possible to upload images and share them with the rest of the world through its push and pull system. Here we can also find already built images that we can use as a basis to build our own.

Once researchers have developed the Dockerfile of their experiment, they can build the corresponding image and push it to DockerHub. When researchers work in an environment other than the local one, it is enough to pull the image from DockerHub to have access to their work environment. Within our developed workflow, this repository can be used as a download point for images. When researchers start working on an HPC cluster, they just need to pull the image from this repository and run the corresponding container to have a functional environment to train models in multiGPU settings.

DockerHub can also be useful for other researchers as they have direct access to the work environment of their colleagues, which facilitates both the replication of experiments and the development of new ones based on already developed computing environments.

In addition to DockerHub, with udocker researchers can use any other private repository to pull images from. In this work we focus on DockerHub, but the extension to a different repository is trivial. 

\subsection{Dockerfile Good Practices}
\label{sec:3:4}

The Dockerfile is composed of code so, as in this kind of project, it is important to follow best practices to  produce clean, understandable and functional code. Docker includes in its official documentation the best practices for Dockerfile\footnote{\url{https://docs.docker.com/develop/develop-images/dockerfile\_best-practices/}}. If these recommended practices are not followed, we will generate a poor quality Dockerfile. According to the nomenclature of configuration smells \cite{sharma_does_2016}, the instructions that contribute to deteriorating this quality are known as Dockerfile smells. 

In \cite{wu_characterizing_2020} these smells are classified into two main groups: DL-smells and SC-smells. The former is related to the violation of Docker best practices and contributes to building failures, increased build latency and security issues. The latter is related to best practices of shell scripting, a fundamental pillar of Dockerfile, its violation can cause strange behaviors and failures under certain conditions.

In \cite{wu_characterizing_2020} a study of the number of projects affected by these smells is carried out. It concludes that $84\%$ of the Dockerfiles studied have some kind of smell. In the case of our developed workflow, it is important to control this mainly for two reasons. The first is to make our work environment understandable and ease maintainability. The second is to generate a clean and light image, since in our workflow we transmit the image over the network (DockerHub) between our local environment and the HPC clusters.

Some specified smells contribute to increasing the size of the images, these are known as temporary file smells \cite{lu_empirical_2019}. We have already seen that each command in the Dockerfile generates a layer that once stacked is read-only. Therefore, Docker images can be considered as append-only. If in one of these layers we generate a temporary file and subsequently we delete it in another, the space is not freed, only its reference is removed (the original layer with the temporary file is read-only and has its own filesystem). Therefore, if we want to delete temporary files, we must do it in the same layer in which they are generated, that is, the same Dockerfile command that generates those files is the one that must also delete them. 

There are many ways to avoid these smells: using \texttt{ADD} instead of \texttt{COPY} when we want to decompress a local file in our image, avoiding the use of temporary files, or always ending the shell command with an \texttt{rm}. It is important to take all this into account to reduce the time elapsed on push and pulls. 

Some tools help us detect and correct Dockerfile smells, one of these is Haskell Dockerfile Linter. This parses the Dockerfile into an abstract syntax tree and applies rules to see if best practices are followed, if not, it detects the smell and provides the solution to the user. There is an online version\footnote{\url{https://hadolint.github.io/hadolint/}} in which the Dockerfile is pasted and returns the smells and their solutions. Researchers can use this tool to remove smells, especially those that contribute to large image size. This can be done once the Dockerfile has been developed, so the image pushed to Dockerhub is already the optimized one, preferably through software quality assurance tools \cite{orviz_fernandez_software_2020}.

\subsection{udocker}
\label{sec:3:5}

As we have already seen, the dependence on administrator intervention or recent Linux kernel features of most of the containerization tools made udocker a perfect candidate for virtualization: it only depends on Python, can be executed under different execution modes and can be installed with not more than the user's privileges. Hereby, udocker is in charge of providing us with the \emph{virtualized} runtime environment. 

\begin{figure}[ht]
    \centering
    \includegraphics[scale=0.1]{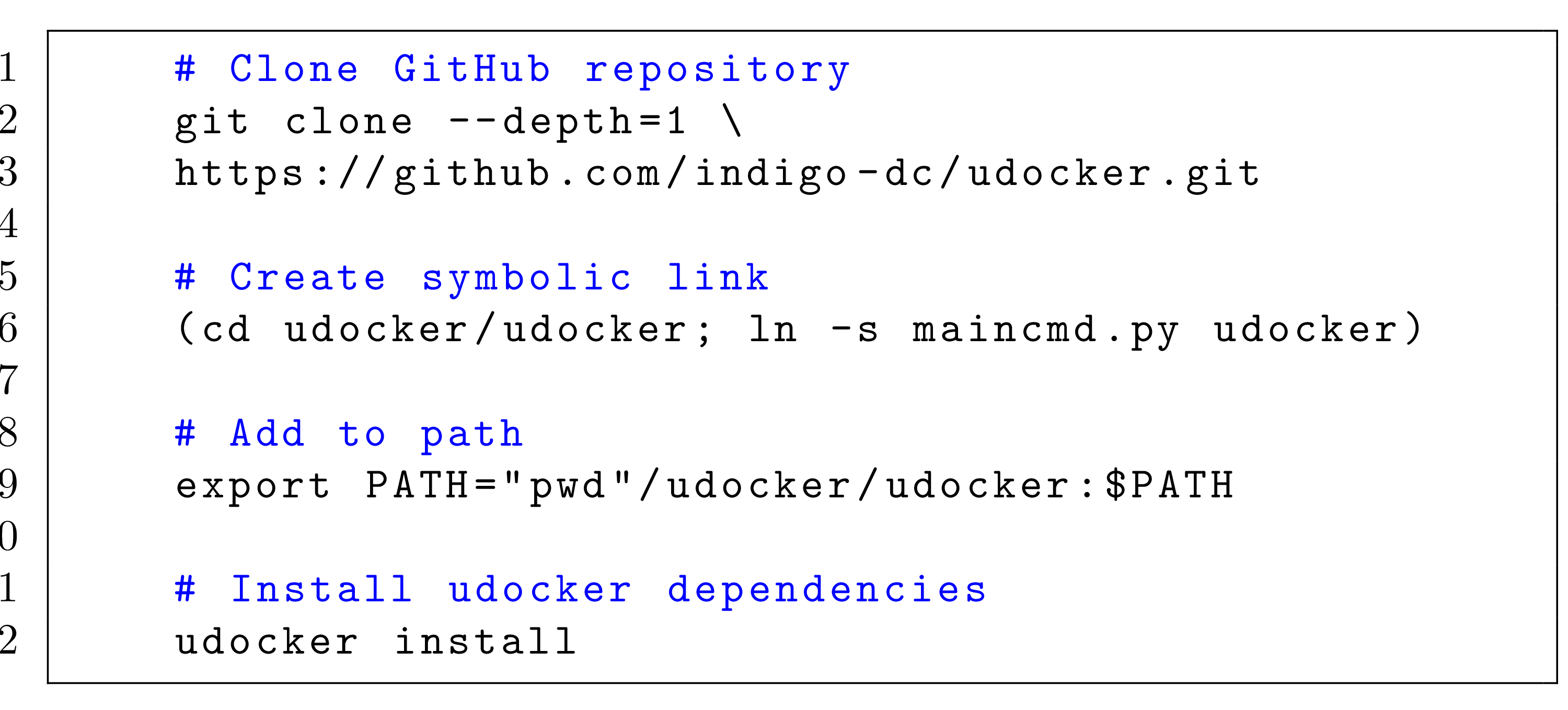}
    \caption{Bash script to install udocker via GitHub}
    \label{fig:3}
\end{figure}

To work with udocker, researchers must install it on the cluster where they will launch their experiments. The main requirement is the availability of Python, which should be straightforward given that this programming language is currently the most popular according to the PopularitY of Programming Language (PYPL) index\footnote{\url{https://pypl.github.io/PYPL.html}}, especially in data science and scientific computing fields. It also requires some basic tools like \texttt{tar}, \texttt{find} and \texttt{curl} which are usually available in such environments. udocker can be installed in multiple ways such as GitHub, tarballs or pip. Figure \ref{fig:3} shows an example of installation via GitHub. This code downloads and executes the \texttt{udocker} Python script and installs the required dependencies. Once this is done, researchers have an executable that can use to run all the udocker commands.

\begin{figure}[ht]
    \centering
    \includegraphics[scale=0.11]{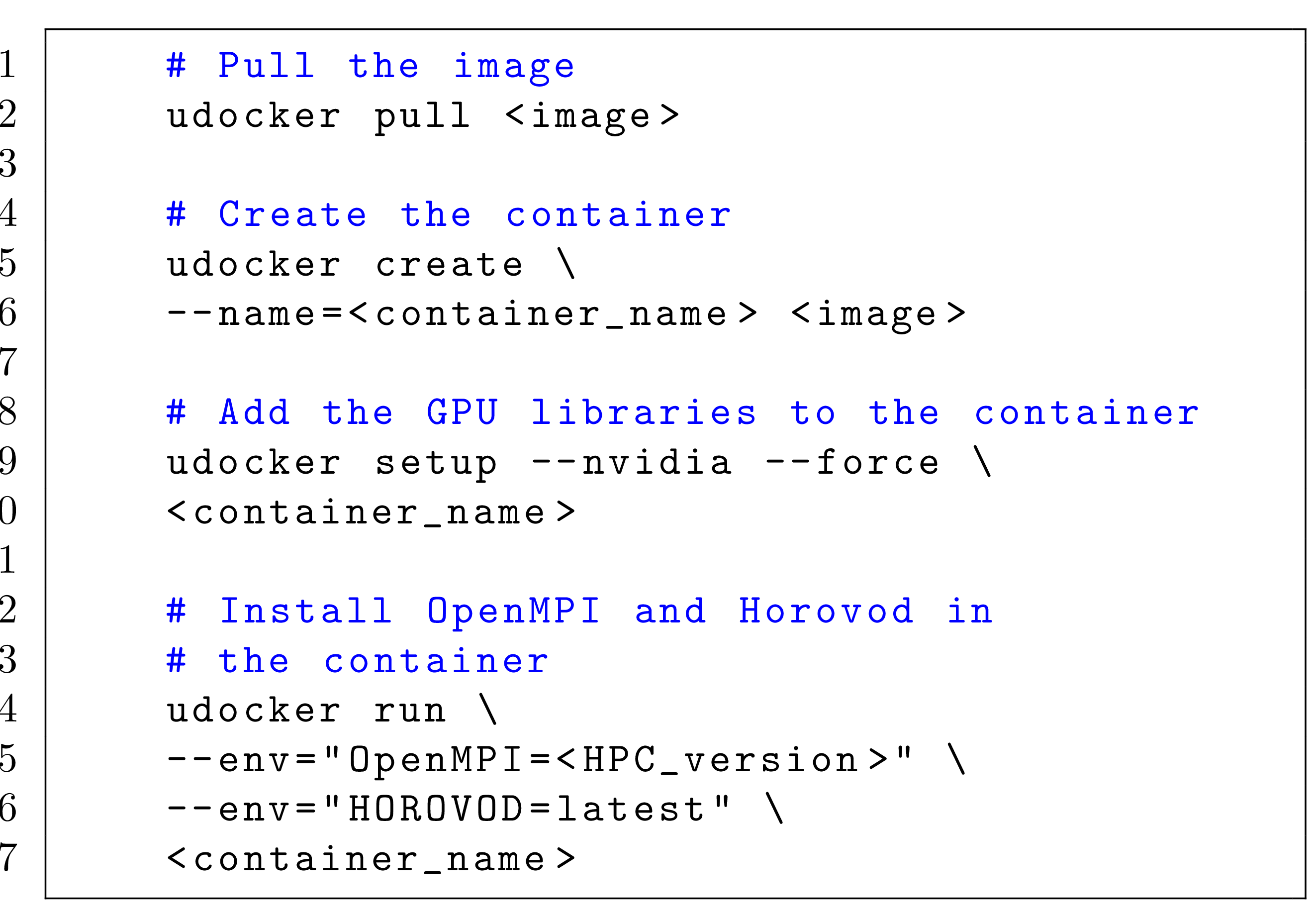}
    \caption{Bash script to create and configure an udocker container}
    \label{fig:4}
\end{figure}

To run a container, udocker uses a syntax similar to Docker. Suppose researchers have developed a Dockerfile under this workflow, they have uploaded the image to DockerHub and installed udocker on the cluster. To create a container following, for example, the approach of Section \ref{sec:3:2:3}, they just need to execute the code of Fig.\ref{fig:4}. First, the
image is pulled from DockerHub to the cluster and the corresponding container is created. Then with the command \texttt{udocker --setup nvidia} the necessary host libraries are added to the container so that it has access to the GPUs of the node. Finally, the container is executed, specifying the version of OpenMPI (the same as in the cluster) and Horovod to install. Once finished, a container ready to distribute the training of deep learning models in the cluster is configured.

To distribute the training, OpenMPI submits the processes as containers, so in each node a different copy of the container can be found. These containers can communicate between each other, allowing for parallel computation. Horovod coordinates distributed work via MPI, so in this case it is straightforward to communicate the different deployed containers as workers, allowing for the parallelization of the training. It is possible to use the \texttt{horovodrun} API or directly OpenMPI through \texttt{mpirun} and \texttt{mpiexec}. In the official Horovod documentation\footnote{\url{https://horovod.readthedocs.io/en/stable/}} researchers can find how to execute their container through these two approaches. This is straightforward and requires no adjustments. We recommend using the OpenMPI approach directly to have more control over the execution of the processes.

\begin{figure}[ht]
    \centering
    \includegraphics[scale=0.11]{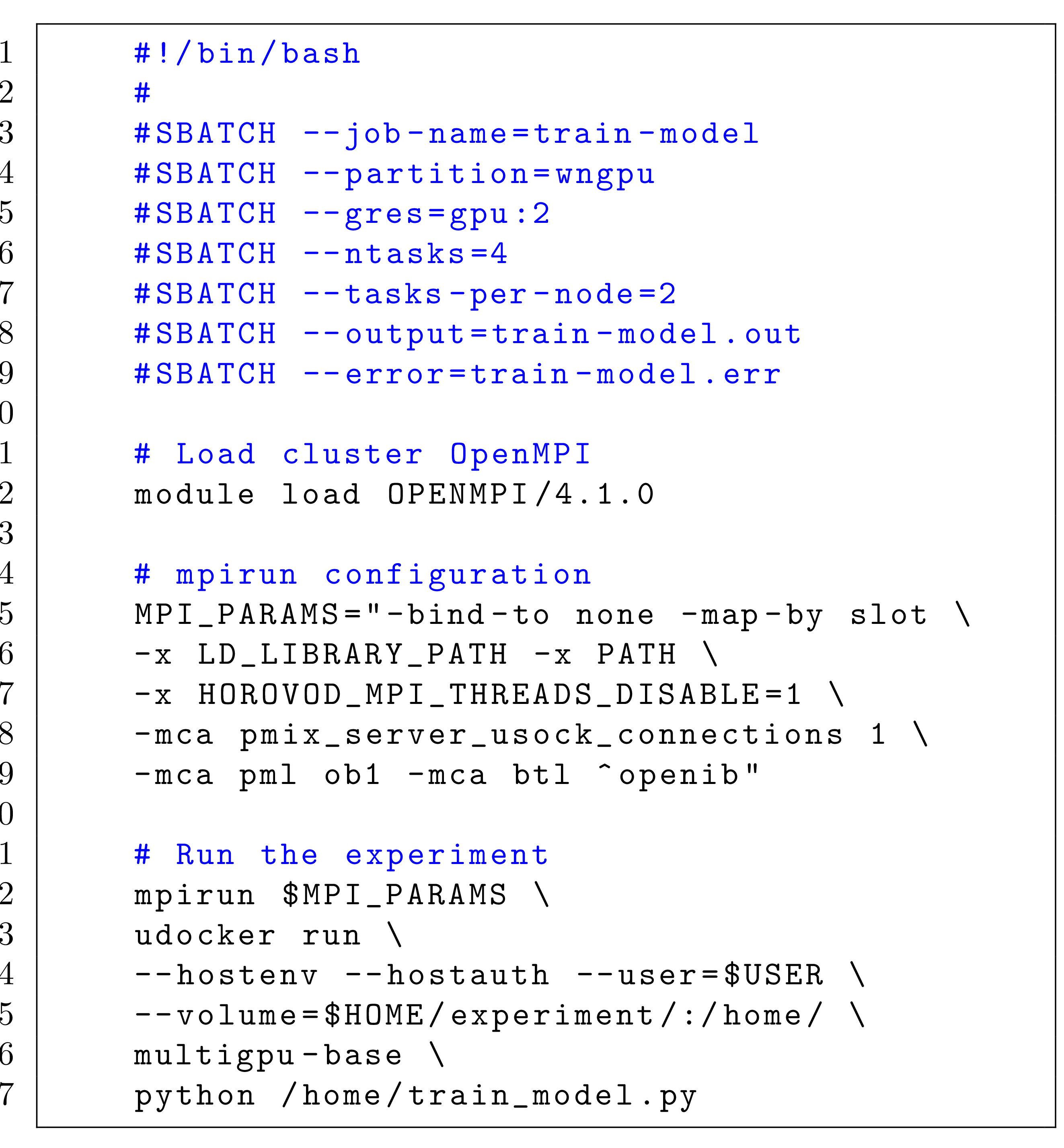}
    \caption{Example of bash code to run an experiment through an udocker container in an HPC cluster using Slurm}
    \label{fig:5}
\end{figure}

Figure \ref{fig:5} shows an example of code to run an experiment in an HPC cluster via udocker. In this case, the execution of a script is being distributed across 2 nodes with 2 GPUs each (4 GPUs in total). Given that this is executed through a job in a Slurm batch system, the \texttt{SBATCH} directives must be defined. These depend on the cluster configuration and the batch system in use, although they do not usually vary much. Then the cluster's OpenMPI is loaded and the \texttt{mpirun} parameters defined (we took as reference the Horovod documentation mentioned before). By default, all communication between GPUs is carried out through NCCL and the rest through MPI, taking advantage of low latency communications through specialized interconnects such as Infiniband. 

Finally, \texttt{udocker run} is executed through \texttt{mpirun} which is responsible for distributing the container in the selected nodes and communicating all the processes. udocker takes as input the information related to the researcher's user in the cluster, the path to the volume to mount in the container and the command to execute. If this command runs a Python script with a Horovod adapted deep learning training, it will be automatically distributed across the selected nodes.

The conjunction between DockerHub and udocker provides researchers with an easily shareable workflow between their local and HPC environment in which they will run their experiments. They can develop their image locally with Docker, upload it to DockerHub and access it through udocker on any HPC cluster. They could also run their experiment with Docker on a small scale in their local environment to verify that everything works correctly and once it is ready, upload it to DockerHub to pull it to the HPC cluster and run it. Once researchers own a ready-to-use image of their working environment, they can work across multiple clusters without the need for specific installations or root access, they just need to pull the image from DockerHub to the respective cluster and start working with their custom environment. 

This methodology provides researchers with a dynamic workflow between local environments and HPC clusters, avoiding complex installations based on the environment used (which would be the case if containerization is not used) and providing them with a closed environment that facilitates the deployment of experiments in a cluster-agnostic way.

\section{Evaluation}
\label{sec:4}

In this section we present the developed experiments to assess the characteristics of the workflow presented in this work: cluster-agnosticism and scaling efficiency. For the former, different scientific workflows are executed in different clusters, using the same starting image as well as the same code. For the latter, different deep learning models are trained in various multiGPU distributed environments with the aim to study its scalability. The code\footnote{\url{https://github.com/IFCA/workflow-DL-HPC}} (including Dockerfiles) and the Docker images\footnote{\url{https://hub.docker.com/r/gonzabad/multigpu-horovod}} for these experiments are publicly available.

\subsection{TensorFlow Benchmark}
\label{sec:4:1}

Horovod includes TensorFlow Benchmark\footnote{\url{https://github.com/horovod/horovod/tree/master/examples/tensorflow2}}, an implementation of SOTA deep learning models used to test training performance both in CPU and GPU environments. This benchmark allows training models like ResNet \cite{he_deep_2016}, InceptionV3 \cite{szegedy_going_2015} and DenseNet \cite{huang_densely_2017} on synthetic and real data. Particularly, we have trained ResN\-et50, ResNet101 and InceptionV3 on synthetic data to avoid disk I/O uncertainty. Table \ref{tab:1} shows key characteristics of the models employed.

\begin{table}[ht]
    \centering
    \begin{tabular}{lll}
    Model & Number of Parameters & Batch size per GPU \\\hline
    InceptionV3 & 23,851,784 & 256 \\
    ResNet50 & 25,636,712 & 256 \\
    ResNet101 & 44,707,176 & 128
    \end{tabular}
    \caption{\label{tab:1}Characteristics of the models trained in the TensorFlow Benchmark experiment}
\end{table}

The objective of this experiment is to verify that the containerization environment scales correctly with respect to the number of GPUs and the number of nodes in comparison with a non-containerized scenario. For this, we train these models in a containerized and non-containerized (native) environments. The former is implemented through the developed workflow, while the latter is not.

This experiment requires a basic environment with a GPU based Tensorflow and Horovod installation. For the workflow test we have prepared a basic image\footnote{\url{https://hub.docker.com/layers/gonzabad/multigpu-horovod/base/images/sha256-2b04fcdb1f7727abb4ba67fcaa12aefe36d2606f1a25ef97bcab8be6ae7e7c00?context=explore}} satisfying these requisites (OpenMPI and Horovod versions issue is addressed by the technique described in Section \ref{sec:3:2:3}). This image can be used as a basis for building more complex environments. udocker allows the use of different execution modes for running the containers, we have opted for the default (P1) which uses the PRoot model with SECCOMP filtering (we refer the reader to \cite{gomes_enabling_2018} for more details). For the native environment we had to depend on software already installed on the HPC cluster, in addition to manually performed installations. For this last part we relied on Anaconda\footnote{\url{https://www.anaconda.com/}}, one of the most popular set of packages and tools among the data science community. This installation is generally more complex than our proposed method, since we heavily depend on the versions of the software installed in the cluster. For example, the version of cudatoolkit available through Anaconda does not include the NVIDIA CUDA compiler (NVCC), so users must install it manually from the NVIDIA repositories, taking into account the actual driver and CUDA version of the cluster. Dependency management is also a complex task, requiring specific configurations to avoid incompatibilities between the different components of the NVIDIA toolkit, as well as requiring the configuration of additional libraries for MPI implementation. Horovod provides documentation for the proper configuration in a local environment\footnote{\url{https://horovod.readthedocs.io/en/stable/conda\_include.html}}.

\subsection{Empirical Statistical Downscaling}
\label{sec:4:2}

Climate modelling plays a critical role in decision-making, in areas ranging from agriculture and health to tourism and economy. This information is generated by Global Circulation Models (GCMs), mathematical models that compute the climate system components and their interactions. These models are highly complex, so the generated predictions suffer from a coarse spatial resolution. Sometimes it is necessary for this information to be available at a higher-resolution in order to make decisions on local scale.  Empirical statistical downscaling (ESD, \cite{maraun_statistical_2018}) methods aim to learn a statistical relationship between a set of low-resolution variables and a high-resolution variable of interest. When these variables lean on observational records, we work under the \textit{perfect-prognosis} approach (PP-ESD).

Given the success of deep learning in fields like computer vision, these kinds of techniques have recently been successfully applied to PP-ESD \cite{vandal_deepsd_2017,bano-medina_configuration_2020,sun_statistical_2021}. In particular, convolutional neural networks have been positioned as a promising approach given the ability to model relationships where the data present a spatial structure.

We develop a PP-ESD experiment following the framework presented in \cite{bano-medina_configuration_2020}. We downscale precipitation over the region of North America using deep learning techniques. As predictor we use five thermodynamical variables at four different vertical levels of the ERA-Interim reanalysis dataset \cite{dee_era-interim_2011} ($2^\circ$ resolution) and as predictand the daily accumulated precipitation from EWEMBI \cite{lange_earth2observe_2019} ($0.5^\circ$ resolution). We train the model in the period spanned by the years 1979-2002. ERA-Interim data can be downloaded from the ECMWF website\footnote{\url{https://www.ecmwf.int/}} and EWEMBI dataset is available at ISIMIP\footnote{\url{https://www.isimip.org/}}.

We have developed a custom model exclusively composed of convolutional layers, what makes it fully-convolutional. Our objective with this experiment is not to improve the accuracy of previous models, but to prove that the workflow possess benefits for a real scientific case. Table \ref{tab:2} shows details of the model trained.

\begin{table}[ht]
    \centering
    \begin{tabular}{lllllll}
    \begin{tabular}{@{}c@{}}Number of \\ Parameters\end{tabular} & Input Size & Output Size & \begin{tabular}{@{}c@{}}Batch Size \\ per GPU\end{tabular} & Optimizer & Learning Rate \\\hline
    1,615,671 & (30, 53, 20) & (117, 209, 3) & 32 & Adam & 0.0001
    \end{tabular}
    \caption{\label{tab:2}Characteristics of the fully-convolutional model trained in the Statistical Downscaling experiment}
\end{table}

In this case we work with R and Python. The former is used to pre-process and post-process the data, while the latter is responsible to distribute the training of the model (via Horovod). Training has been implemented with Keras, a high level API of TensorFlow. To adapt it to Horovod we only needed to add a few lines of code. This has been done following the official Horovod documentation. In this experiment the use of R is also justified by climate4R \cite{iturbide_r-based_2019}, a bundle of R packages for access, pre-process, post-process and visualization of climate data. In this case, to configure the environment we just need to add to the Dockerfile the commands required to install R and climate4R. As a result, we get a ready-to-use image satisfying the requirements of our experiment.

The goal of this experiment is to show how the workflow can help in moving experiments between clusters while maintaining its scalability in multiGPU environments. This experiment was executed in a different cluster from the previous one, however it was not necessary to apply substantial changes to the workflow. The image built for this experiment\footnote{\url{https://hub.docker.com/layers/gonzabad/multigpu-horovod/downscaling/images/sha256-d695141efc6677e0ac38e6702c5a0c580ed8bc7613a6e5063e609c3e83a738b1?context=explore}} was developed using as basis the image from the TensorFlow Benchmark experiment, solely adding the software required for this second experiment. The scheduling of the jobs has been done in the same way in both experiments (following the indications of section \ref{sec:3:5}), modifying only the \texttt{SBATCH} directives since they depend on specific configurations of each cluster. This has been possible thanks to the workflow developed and its cluster-agnosticism feature.

\section{Cluster Setup}
\label{sec:5}

The experiments have been executed on two different computing clusters: one at the Spanish National Research Council (CSIC) deep learning computing infrastructure located at the Institute of Physics of Cantabria (IFCA, CSIC--UC) in Spain and ForHLR2 at the Karlsruhe Institute of Technology (KIT) in Germany. Within these clusters the nodes with GPUs are the ones accessed. Table \ref{tab:3} shows the specifications of each of these two clusters.

\begin{table}[ht]
\begin{tabular}{|c|c|c|}
\hline
\multicolumn{1}{|l|}{}                           & \textbf{CSIC DL Infrastructure}                                                                                       & \textbf{ForHLR2}                                                                    \\ \hline
\multicolumn{1}{|c|}{\textbf{CPU}}              & \begin{tabular}[c]{@{}c@{}}Intel(R) Xeon(R) Gold\\ 6230 CPU @ 2.10GH\end{tabular}                                     & \begin{tabular}[c]{@{}c@{}}Intel Xeon E7-4830 v3\\ processors @ 2.10GH\end{tabular} \\ \hline
\multicolumn{1}{|c|}{\textbf{Memory}}           & 64 GB                                                                                                                 & 1 TB                                                                                \\ \hline
\multicolumn{1}{|c|}{\textbf{GPU}}              & NVIDIA Tesla V100                                                                                                     & \begin{tabular}[c]{@{}c@{}}NVIDIA GeForce\\ GTX980 Ti\end{tabular}                  \\ \hline
\multicolumn{1}{|c|}{\textbf{GPU Memory}}       & 32 GB                                                                                                                 & 6 GB                                                                                \\ \hline
\multicolumn{1}{|c|}{\textbf{GPUs per node}}    & 2                                                                                                                     & 4                                                                                   \\ \hline
\multicolumn{1}{|c|}{\textbf{GPU nodes}}        & 20                                                                                                                    & 21                                                                                  \\ \hline
\multicolumn{1}{|c|}{\textbf{Connection}}       & \begin{tabular}[c]{@{}c@{}}InfiniBand EDR NVIDIA Mellanox \\ ConnectX-5 cards\end{tabular}                            & \begin{tabular}[c]{@{}c@{}}InfiniBand 4X EDR\\ Interconnect\end{tabular}            \\ \hline
\multicolumn{1}{|c|}{\textbf{Operating system}} & Ubuntu 18.04.4 LTS                                                                                                    & \begin{tabular}[c]{@{}c@{}}Red Hat Enterprise \\ Linux (RHEL) 7.x\end{tabular}      \\ \hline
\multicolumn{1}{|c|}{\textbf{Storage}}          & \begin{tabular}[c]{@{}c@{}}General Parallel File System\\ (GPFS \cite{schmuck2002gpfs})\end{tabular} & Lustre \cite{braam2019lustre}                                      \\ \hline
\end{tabular}
\caption{Specifications (rows) for the two different clusters (columns) accessed for the different experiments}
\label{tab:3}
\end{table}

The experiments of Section \ref{sec:4:1} have been executed at IFCA while those of Section \ref{sec:4:2} at ForHLR2.

\section{Results}
\label{sec:6}

\subsection{TensorFlow Benchmark}
\label{sec:6:1}

For this experiment, we have trained the three models mentioned in Section \ref{sec:4:1} in configurations of 1, 2, 3, 4, 5 and 6 GPUs. Training sessions consisted of 10 warm-up iterations (not counting towards measurements) and 10 iterations from which the final measures were obtained. Each iteration processed 10 batches of data, each with the number of synthetic samples specified in Table \ref{tab:1}.

\begin{figure}[ht]
    \centering
    \includegraphics[scale=0.38]{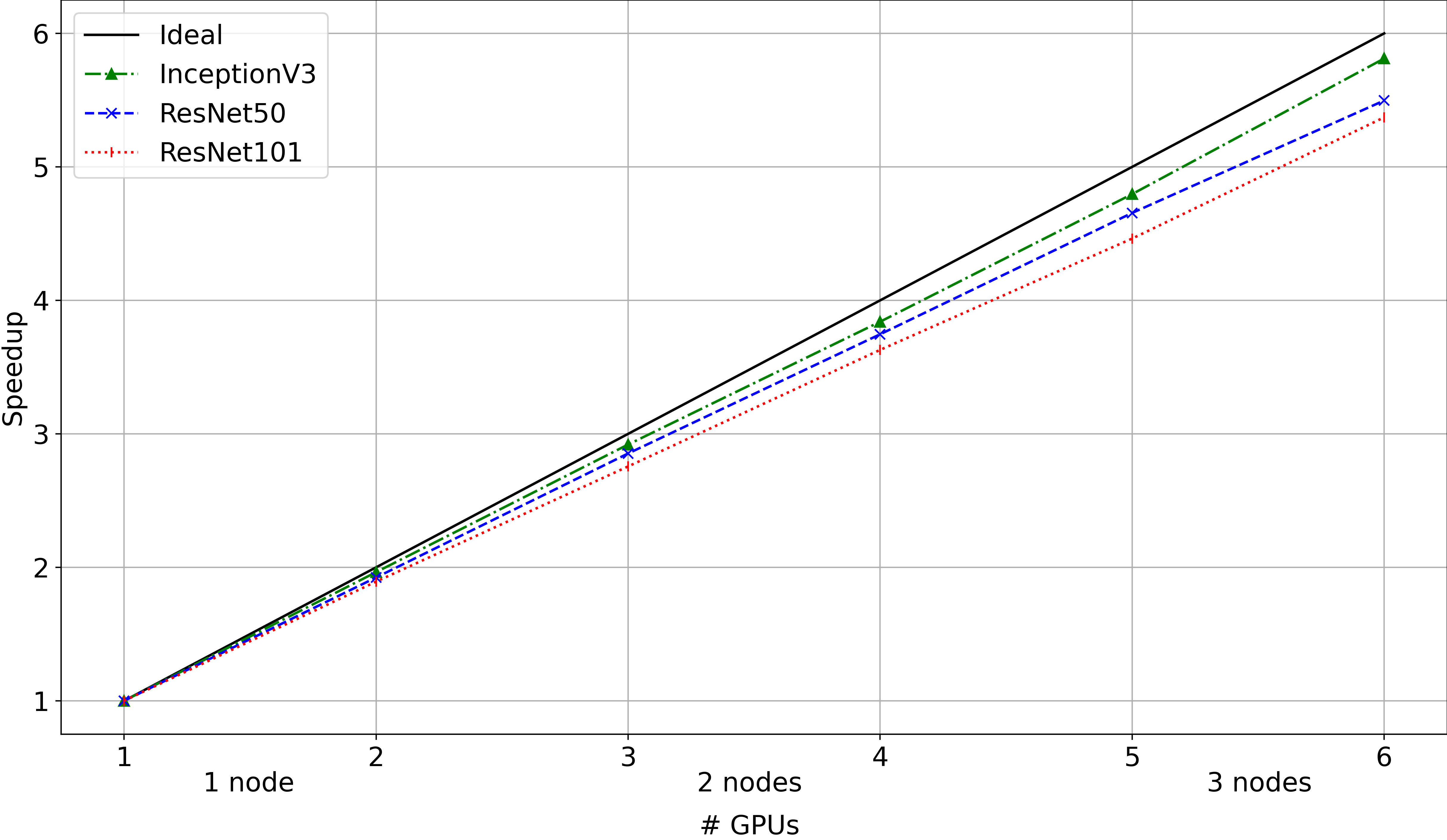}
    \caption{Speedup plot for the TensorFlow Benchmark experiment using udocker over the different GPU configurations}
    \label{fig:6}
\end{figure}

Figure \ref{fig:6} shows the speedup achieved for the six configurations concerning the number of images processed per second when training the models under the udocker workflow. For one and two nodes the speedup in training is very close to the ideal. For the configurations running on three nodes, although not ideal, acceptable results are achieved. Among the models, ResNet101 is the one that scales the worst, followed by the ResNet50 and InceptionV3. This order coincides with the number of parameters, the more parameters the model has, the lower the speedup.

\begin{figure}[ht]
    \centering
    \includegraphics[scale=0.38]{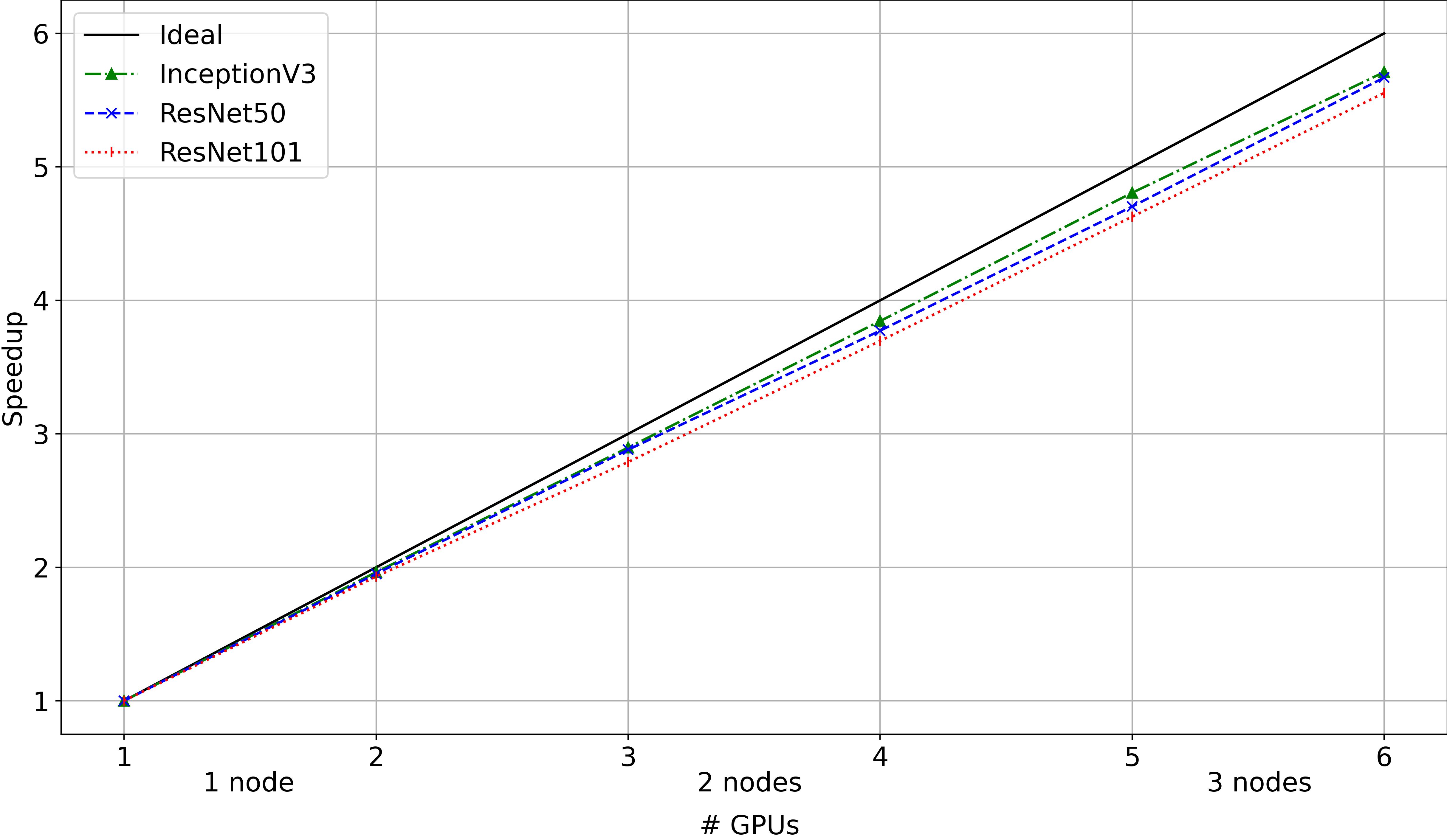}
    \caption{Speedup plot for the TensorFlow Benchmark experiment using the native environment over the different GPU configurations}
    \label{fig:7}
\end{figure}

Figure \ref{fig:7} shows the results for the same experiment but in the native environment. These depict a similar trend in the scalability with respect to the workflow experiment, however the speedup seems to be a little bit closer to the ideal one. In addition, the difference in speedup between the three models is smaller, although it can still be appreciated.

\begin{figure}[ht]
    \centering
    \includegraphics[scale=0.38]{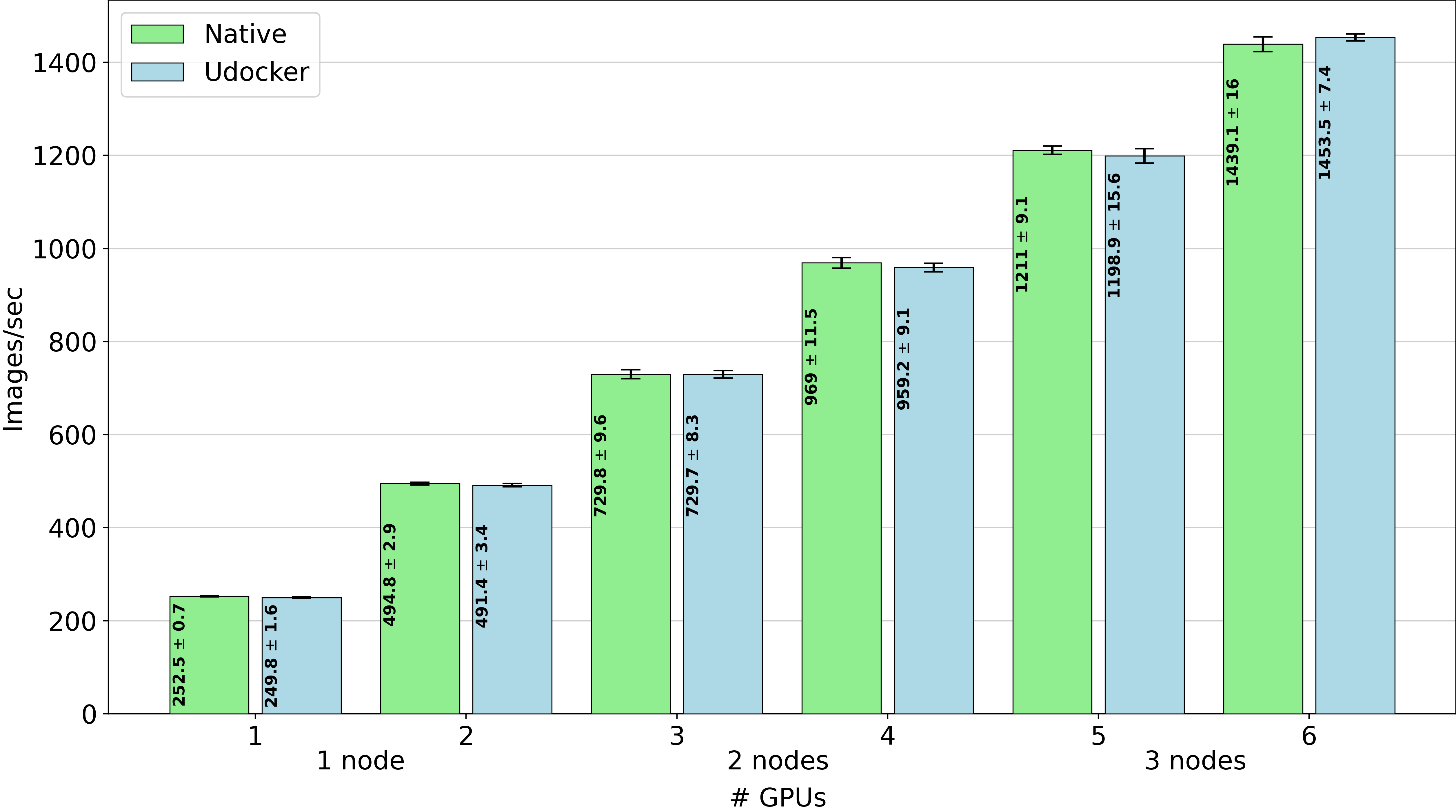}
    \caption{Comparison of images processed per second when training the InceptionV3 model. The values represented correspond to the mean across different iterations. The 95$\%$ confidence interval can be found within each bar.}
    \label{fig:8}
\end{figure}

\begin{figure}[ht]
    \centering
    \includegraphics[scale=0.38]{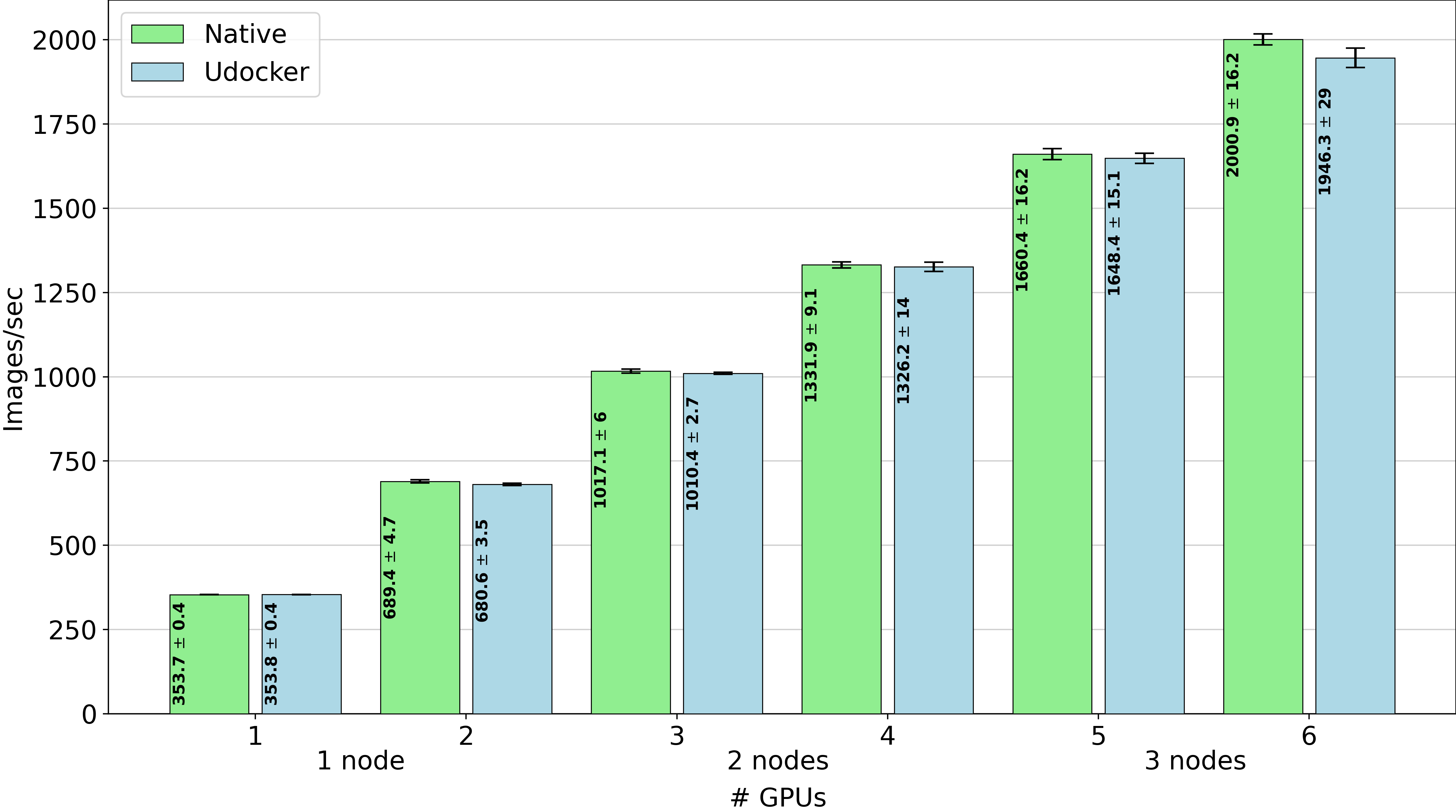}
    \caption{Comparison of images processed per second when training the ResNet50 model. The values represented correspond to the mean across different iterations. The 95$\%$ confidence interval can be found within each bar.}
    \label{fig:9}
\end{figure}

\begin{figure}[ht]
    \centering
    \includegraphics[scale=0.38]{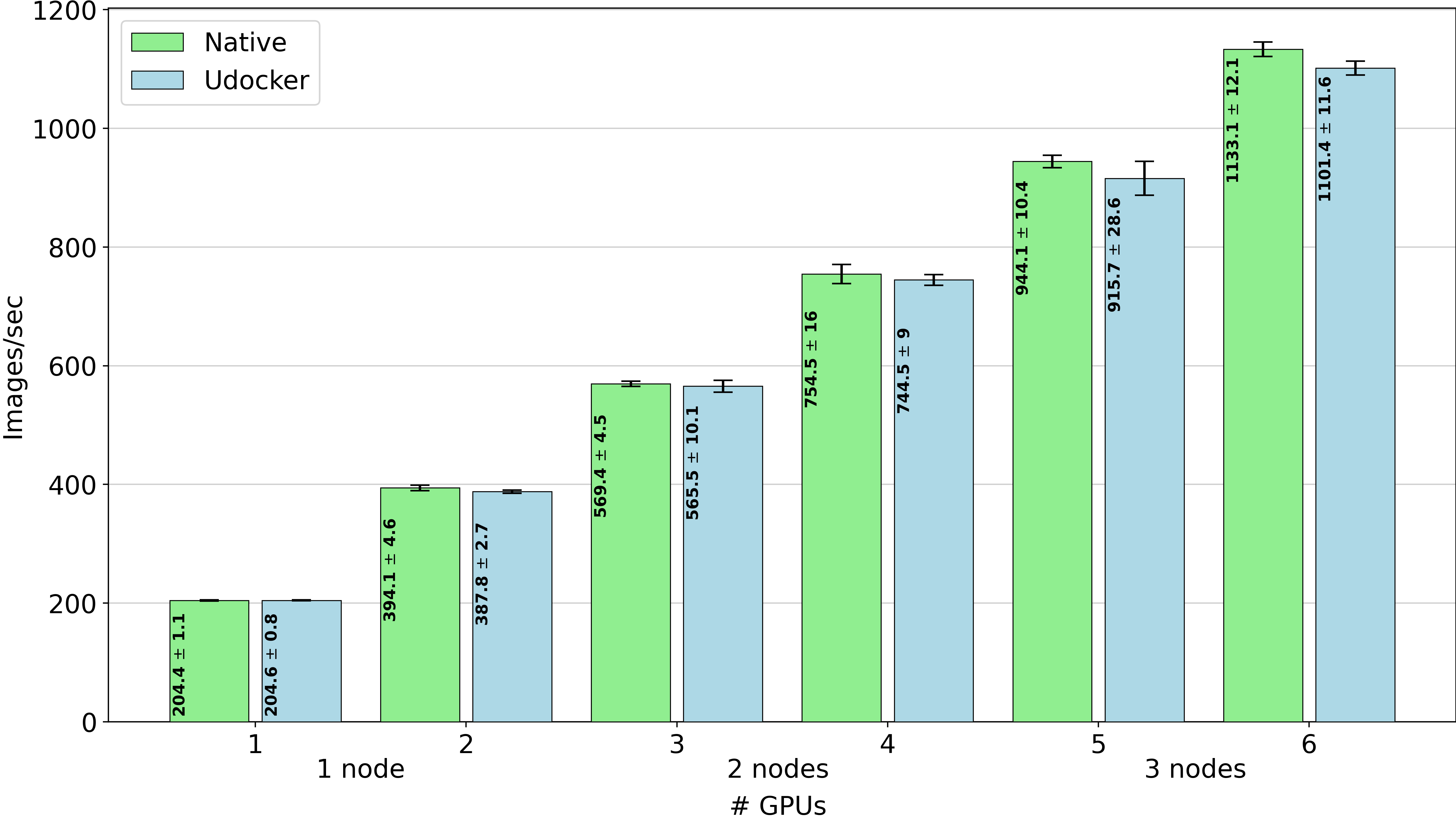}
    \caption{Comparison of images processed per second when training the ResNet101 model. The values represented correspond to the mean across different iterations. The 95$\%$ confidence interval can be found within each bar.}
    \label{fig:10}
\end{figure}

Finally, Figures \ref{fig:8}, \ref{fig:9} and \ref{fig:10} compare the images processed per second in the different configurations under the udocker workflow and the native environment. Values represented as bars correspond to the mean of images per second across the different iterations. Within each bar it can be seen the 95$\%$ confidence interval for these measurements. Concerning udocker and native environments, no significant differences are observed between models, especially for one and two nodes. For three nodes a maximum difference of $3\%$ of processed images in favor of the native environment is observed, although differences benefiting the udocker workflow can also be found (e.g., InceptionV3).

\subsection{Statistical Downscaling}
\label{sec:6:2}

This experiment has been executed in the ForHLR2 cluster. The GPUs available had 6 GB of memory, which was not enough to train the model for the statistical downscaling task when using a batch size of more than 32 samples. We have avoided this issue by scaling the training to more than one GPU. More specifically, we ran the experiment in configurations of 1, 2, 3 and 4 GPUs accessing a single node. For this, we have deployed the distributed training across GPUs with the udocker workflow. Although all GPUs were located on a single node, we still benefited from the Horovod framework. This has allowed us to work with batch sizes of 32, 64, 96 and 128 samples (1, 2, 3 and 4 GPUs respectively).

\begin{figure}[ht]
    \centering
    \includegraphics[scale=0.38]{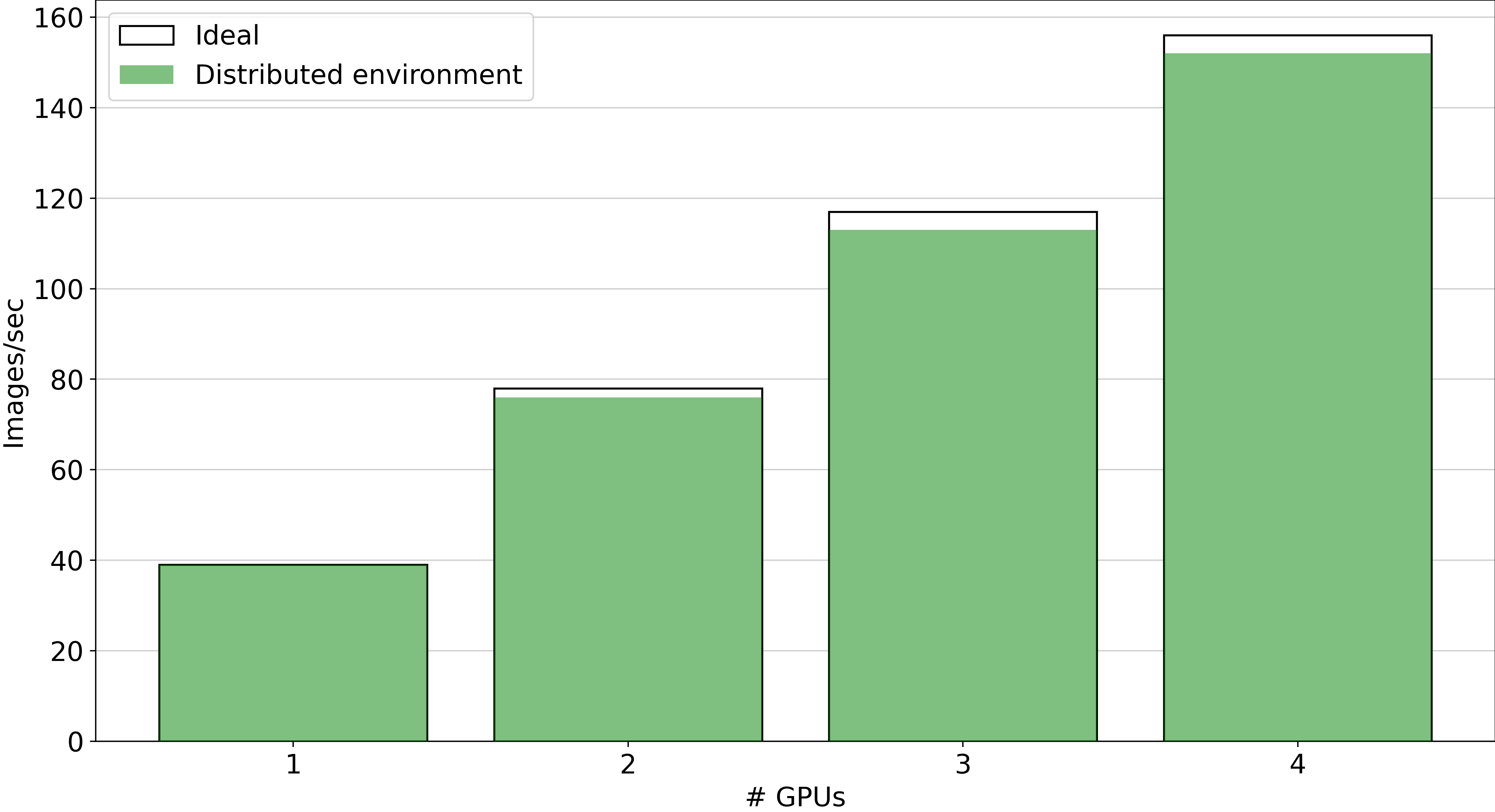}
    \caption{Images processed per second when training the statistical downscaling model using udocker}
    \label{fig:11}
\end{figure}

Figure \ref{fig:11} shows the number of images processed per second concerning the number of GPUs. Results depict an optimal scalability given the similarity between the ideal scenario and the amount of images processed.

\begin{figure}[ht]
    \centering
    \includegraphics[scale=0.38]{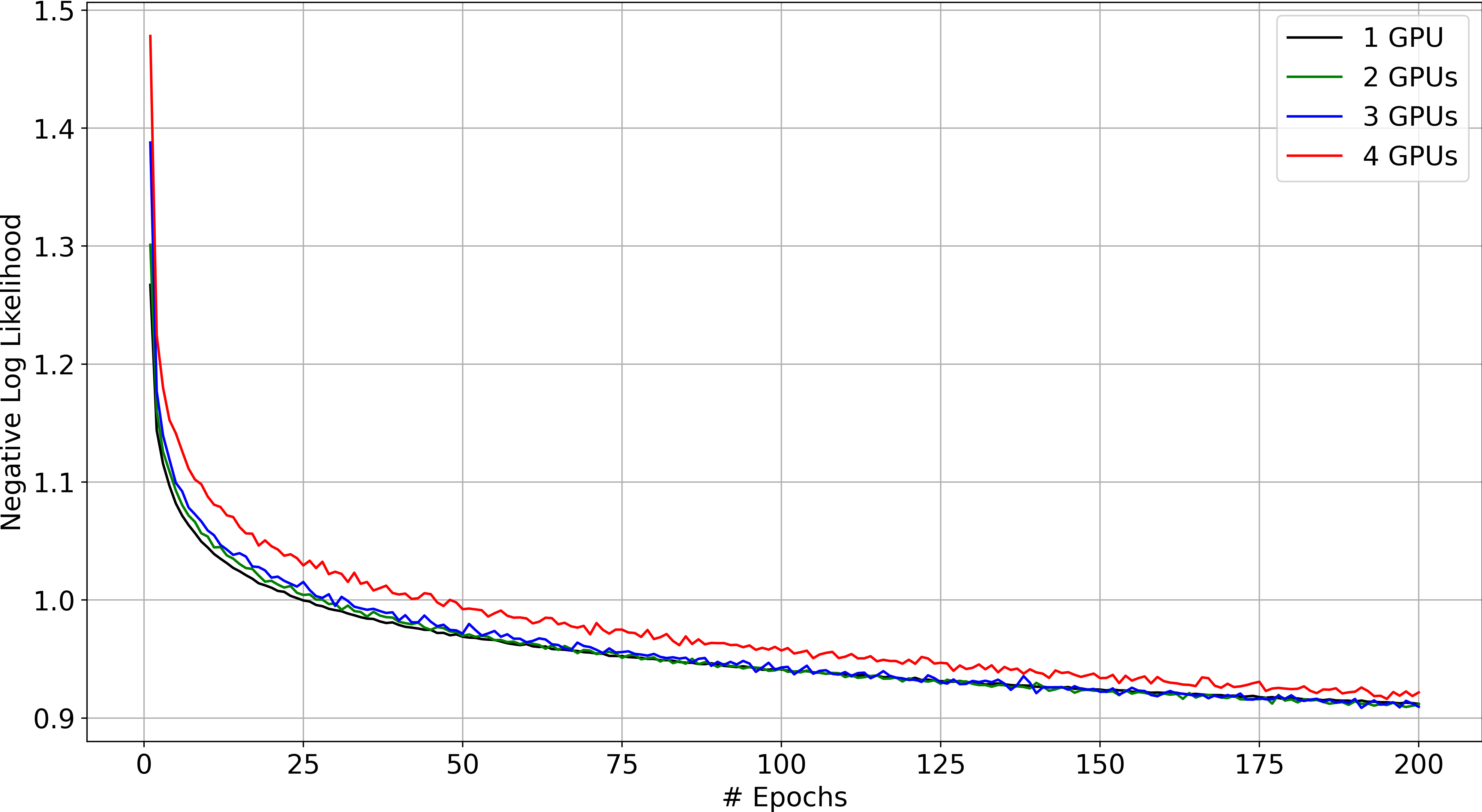}
    \caption{Training Loss (Negative Log Likelihood) evolution during training of the statistical downscaling model}
    \label{fig:12}
\end{figure}

Figure \ref{fig:12} shows the evolution of the loss during training with respect to the number of GPUs. For 1, 2 and 3 GPUs (effective batch size of 32, 64 and 96 respectively) a similar convergence rate can be seen, however for 4 GPUs a slower evolution of the loss is observed, although it ends up reaching the same value.

\section{Discussion}
\label{sec:7}

The presented results show that it is possible to efficiently distribute the training of deep learning models in HPC clusters with the developed workflow. The differences observed in performance between udocker and the native environment are minimal, highlighting an advantage of less than $3\%$ in images processed per second towards the latter. This is an expected effect, since any type of containerization or virtualization can introduce overhead, although  it is usually minimal \cite{torrez_hpc_2019}. No comparisons have been made with models trained with other tools like Singularity, as it has been already seen that their performance is very similar \cite{grupp_benchmarking_2019}.

Regarding the number of GPUs, the results show a small deviation from the ideal scalability. This is expected both in a containerized environment and in a native one since the distribution of the training introduces difficulties (i.e. communication between workers) that slightly damage the speedup. A worse scalability is also observed with respect to the number of parameters of the model, this is due to the increase in the communication time between workers: the greater the model, the more data (parameters) need to communicate over the network. This effect is most noticeable when using udocker, as additional operations are required between workers given the containerization layer.

Although in the case of statistical downscaling the different models achieve a similar loss value, they do it at different speeds. This is an effect generally caused by the use of large batch sizes \cite{DBLP:conf/iclr/KeskarMNST17}. Although advanced techniques exist to mitigate this effect, we followed the Horovod recommendations and used a linear scaling of the learning rate based on the number of GPUs, obtaining an acceptable learning result regardless of the batch size.

The statistical downscaling experiment has been executed in a different cluster than the TensorFlow benchmark, however both the deep learning framework and all the libraries it depends on have been installed following the same image. Also, the scripts to run the jobs have been shared between experiments. Despite significant differences between clusters, such as the GPU model or the operating system on which they run (see Section \ref{sec:5}), there have been no compatibility issues and the experiments have been executed directly, proving the cluster-agnosticism of the workflow.

Installation and configuration of the required software in the clusters for both experiments has been carried out with no need of administrators intervention given the non-root nature of the tools conforming the workflow. The images of both experiments have been developed in a local machine, transferring them to the HPC cluster via Dockerhub. These facts enable users with the autonomy to control their experiments in a more dynamic way, without depending on external individuals.

\section{Conclusions}
\label{sec:8}

This work presents a methodology and workflow based on containers to perform distributed training of deep learning models in HPC clusters. It covers the entire process required for this purpose, from the development of the environment to its distribution and execution. In this way, domain scientists who need to apply deep learning algorithms can take advantage of the computing power available in HPC clusters. We have shown that this method scales well with respect to the number of GPUs and nodes thanks to the use of udocker. We have conducted experiments with a real scientific use case proving that learning is not affected.

Using Docker containers to encapsulate the execution of scientific user applications poses benefits in order to port and execute the workloads in HPC environments. The workflow developed also offers a number of advantages over more conventional ones: improves transferability between different clusters, promotes scientific collaboration and sharing between communities, allows for a more dynamic research cycle given its non-root nature and permits the use of the latest software in a secure way at HPC clusters. With this work we hope to encourage domain scientists to make use of these computing resources in order to further advance their research.

\section*{Acknowledgments}

Authors would like to acknowledge the European Centre for Medium-Range Weather Forecasts (ECMWF) for the ERA-Interim dataset and Potsdam Institute for Climate Impact Research for the EWEMBI dataset.

\bibliography{sn-bibliography}


\end{document}